\documentstyle[11pt,paspconf,epsf]{article}

\begin{document}

\title{Horizontal Branch Stars in the Galactic Stellar Halo}
\author{Andrew C. Layden\altaffilmark{1}}
\affil{U. of Michigan, Dept. of Astronomy, Ann Arbor, MI 48109-1090,
USA \\ layden@astro.lsa.umich.edu}

\altaffiltext{1}{Hubble Fellow.}

\begin{abstract}
I review the properties of the Galaxy's stellar halo, specifically as
traced by horizontal branch stars, but with reference to and in
comparison with other stellar tracers.  I discuss the halo density
profile, the variation in horizontal branch morphology with Galactic
position, and the kinematics of halo stars in several regions of the
Galaxy.  I summarize the horizontal branch star findings and place
them in the context of a currently popular picture of Galactic halo
formation.
\end{abstract}

\keywords{stars, halo}

\section{Introduction}

The surviving ancient stars of the Galactic stellar halo (hereafter,
``halo'') contain information on the locations, motions, and
compositions of the proto-Galactic gas clouds in which they formed.
These stars therefore represent our best means of understanding the
processes occurring during the formation and early evolution of our
Galaxy.  They also provide an important consistency check on studies
of spiral galaxy formation at high redshift and QSO absorption line
systems.
 
The most visible component of the halo is the globular cluster system
(see Harris, this volume).  However, the globulars represent only a
few percent of the mass of the halo, and are subject to tidal
disruption.  Therefore, it is not clear whether they provide a truly
representative picture of the halo.  Furthermore, only $\sim$150
clusters exist, so any statistical arguments based on them have an
intrinsic signal-to-noise limit.
 
Field stars in the halo provide orders of magnitude more objects to
study, and are ``indestructible'' entities.  However, it is more
difficult to determine distances, metallicities, and ages for
individual stars than for clusters of stars.
 
My goal in this paper is to highlight some of the constraints which
halo stars place on our understanding of the formation of the Galactic
halo.  I discuss specifically horizontal branch (HB) stars, but will
link these specific findings with more general ones based on other
stellar tracers of the halo.

HB stars are sub-solar mass stars in the helium core burning phase of
evolution.  Current evolution theory indicates they formed when the
universe was $<$30\% of its present age.  The HB derives its name from
the sequence of stars seen at $M_V \approx +0.6$ in the
color-magnitude diagrams of globular clusters.  HB stars with $0.2 <
(B-V)_0 < 0.4$ are unstable to radial pulsations, and vary with
$V$-band amplitudes of 0.3--1.5 mag and periods of 0.2--0.8 days.
They are called RR Lyrae variables (RRL), whereas the stable HB stars
redward and blueward of the instability strip are referred to as red
and blue HB stars, respectively.
 
Unfortunately, existing survey techniques limit the range of HB star
colors available for study.  The red HB stars ($(B-V)_0 > 0.4$) are
especially difficult to extract from the overwhelming numbers of disk
F- and K-type stars per unit area of the sky.  Likewise, for $(B-V)_0
< 0.0$, it becomes difficult to differentiate efficiently between blue
HB stars and main sequence A- and B-type stars in the disk.  Only for
$0.0 < (B-V)_0 < 0.2$ can blue HB stars be separated efficiently on
the basis of their lower surface gravities.  I will hereafter refer to
these stars as BHB stars.  RRL are usually detected in variable star
surveys by their characteristic light curves.
Thus, our knowledge of the halo field HB star population is derived
from stars spanning $0.0 < (B-V)_0 < 0.4$ mag, only about half the
color extent of the full HB.
 
\section{Structure of the Galactic Halo}

Preston, Shectman \& Beers (1991, PSB) provide the most extensive look
at the distribution of HB stars in the halo.  Their Figure 22 shows
the distribution of stars, both BHB and RRL, from several deep
surveys.  The range of coverage of the Galactic halo enables a
detailed investigation of the density structure of the halo out to a
Galactocentric distance of $R_{GC} \approx 50$ kpc.
 
Figure 1a shows the density of stars as a function of $R_{GC}$. The
PSB BHB stars are shown as crosses.  RRL stars come from several
sources: the PSB compilation is shown as solid squares, the survey of
Hawkins (1984) is shown by stars, the "transit" survey of Wetterer \&
McGraw (1996) is show by open squares, and the local ($d < 2$ kpc)
counts from Layden (1995) are shown by a triangle.  The least-squares
fit to the RRL data is shown by the solid line $[\log \rho = 3.00 -
2.80 \log R_{GC}$], while the fit to the BHB stars is from PSB (slope
= --3.31).

\begin{figure}
\vskip -0.4in
\plottwo{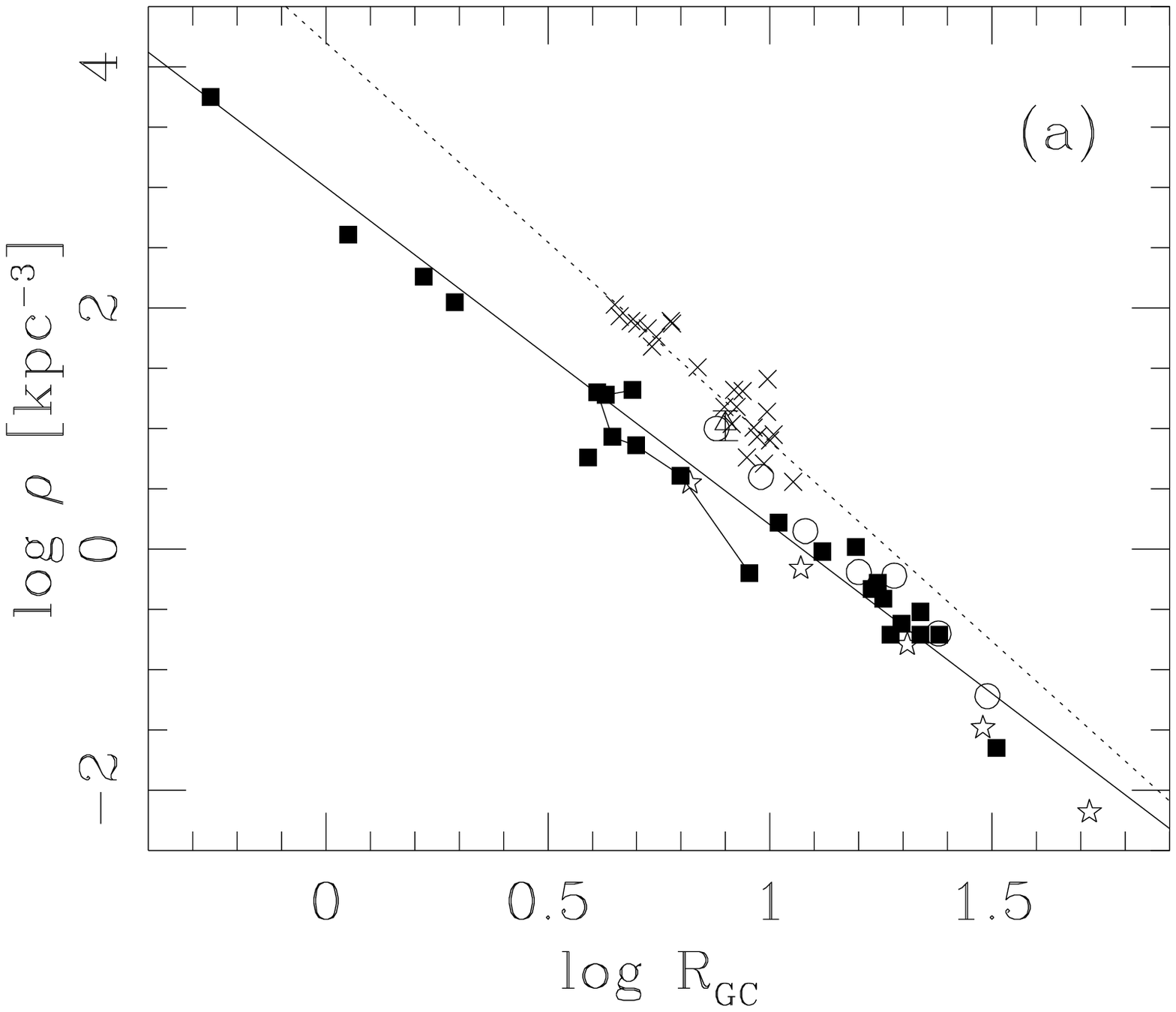}{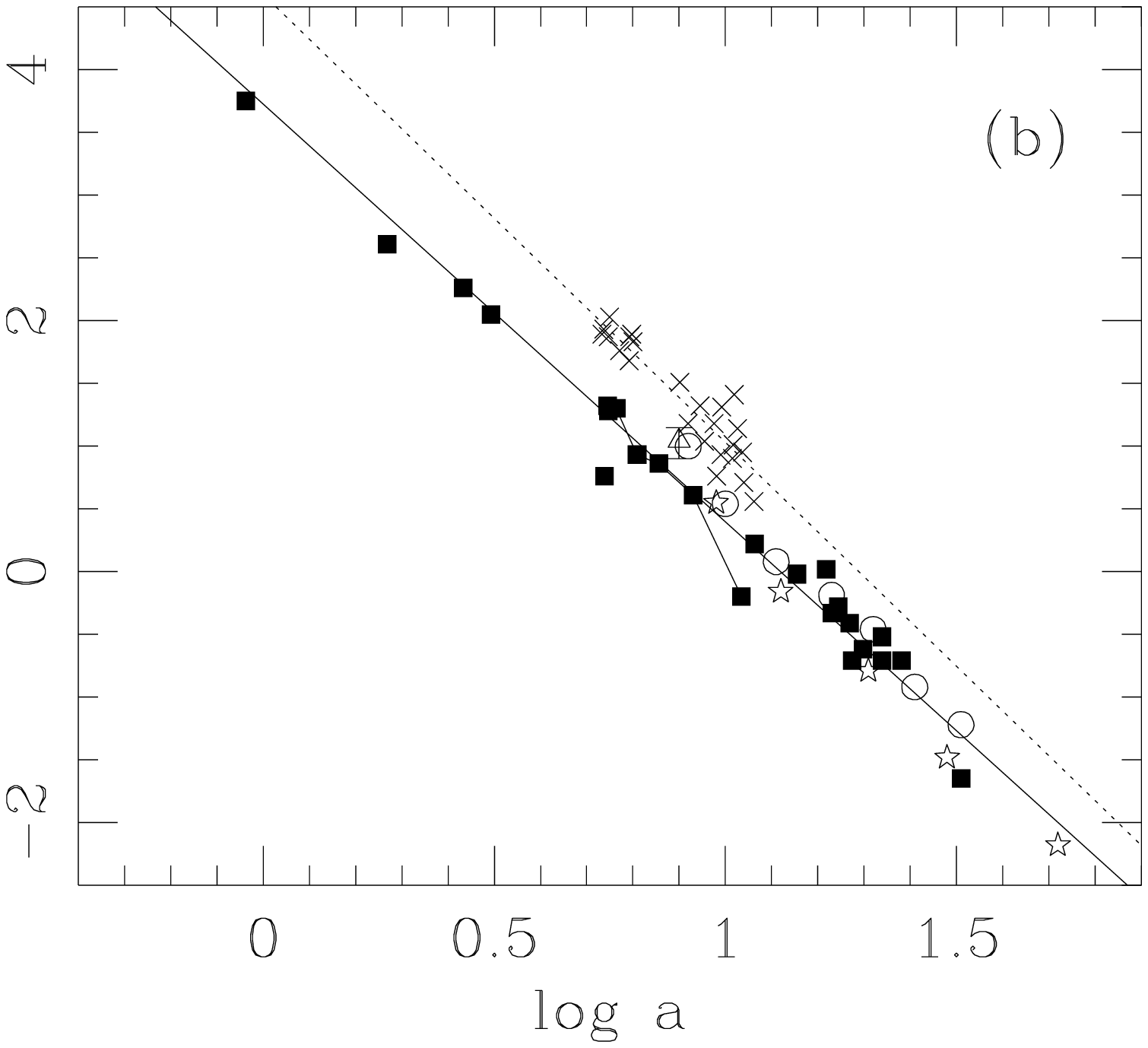}
\vskip -0.25in
\caption{HB star density as a function of (a) Galactocentric 
distance, and (b) ellipsoid semi-major axis.  Symbols are 
described in \S 2.} \label{fig-1}
\end{figure}

Following the work of Kinman (1965), PSB show that the sequence of
points for a particular field which passes through the inner halo
indicates that the isodensity contours in this region must be
flattened towards the plane.  The connected set of points in Figure 1a
shows that the points nearest the Sun (and hence the Galactic plane;
upper end of ``hook'') are overdense compared to more distant points
(farther from the plane).
 
To correct this, PSB adopt a density model in which the ellipsoid axis
ratio ($c/a$) decreases from a central value of 0.5 to 1.0 (spherical)
at 20 kpc.  Figure 1b shows this model applied to the data from Figure
1a.  The ``hook'' has now collapsed on itself, and the scatter about
the best fit is reduced, indicating that this model more accurately
fits the density distribution of halo HB stars than a simple spherical
distribution.  The least-squares fit to the RRL data in Figure 1b is
$[\log \rho = 3.72 - 3.33 \log a$],
where $a$ is the semi-major axis of the ellipsoidal isodensity contours.
The fit to the BHB stars from PSB has a slope of --3.57.
 
It is impressive that a simple model can fit so nicely the run of HB
star densities over 50 kpc in $R_{GC}$ and a factor of $10^6$ in
density.  Furthermore, the lack of a density enhancement inside
$R_{GC} \approx 2$ kpc supports the claim by Minniti (1996)
that the RR Lyrae stars in this region belong to the inner halo, not
the bulge.  The Figure 1 results are in general agreement with the
halo density distribution as traced by the globular clusters ($\rho
\propto R^{-3.5}$, Zinn 1985).

It should also be noted that in their study of HB star densities,
Kinman et al. (1994) interpreted data similar to that shown in Figure
1 as evidence for a two-component halo; a spherical component which
dominates at large distances from the Galactic plane, and a flattened
component which dominates near the Sun.  The density resolution of the
available surveys is not sufficient to distinguish clearly between
these two possibilities.  Clearly, additional RRL density studies of
the inner halo are needed to confirm the presence of the ``hook'' and
to establish and quantify the nature of the density variations.

\section{Horizontal Branch Morphology}

The HB morphology (color distribution of stars along the HB) of a
stellar population at a given abundance tends to be bluer for older
populations, and redder for younger ones.  Estimates of the HB
morphology of the halo field at different positions in the Galaxy can
thus give insights into the relative epochs and durations of halo star
formation in these regions.  However, exact knowlege is hampered by
(1) the possible influence of additional parameters (e.g., He
abundance, stellar rotation, etc.; see Lee et al. 1994) on HB
morphology, and (2) the limited color range over which the HB can be
sampled (see Sec. 1).
 
Figure 1 provides a hint that the field HB morphology may change as a
function of position in the Galaxy.  The fits to the BHB stars have a
steeper slope than for the RRL stars (albeit with large uncertainty),
suggesting that the inner halo has a higher relative density of BHB
stars, i.e. a bluer HB morphology.

In a more intensive study of HB stars in several of these fields,
Kinman et al. (1994) found that the number ratio of BHB to RRL stars
was $>$1 near the Galactic plane, but $\sim$1 at distances $>$5 kpc
from the plane.  Since the metallicity in the two regions was the
same, they suggested that the bluer HB morphology of the flattened
component of their two-component halo indicated that it was older than
the spherical component.
 
However, the relative counting incompleteness between the BHB and RRL
surveys in these fields clouds the issue.  Using BHB stars alone, PSB
showed that the color distribution of BHB stars in their narrow color
window shifts dramatically with direction in the Galaxy (see their
Fig. 1).  Toward the inner Galaxy (angles $A =$ 10--45$^\circ$ from
the Galactic center), the color distribution is strongly peaked to the
blue side of the BHB color window, while at $45 < A < 90^\circ$ and $A
> 90^\circ$ the distribution becomes progressively redder and broader.
The [Fe/H] gradient in the halo is small, and in the sense of lower
[Fe/H] at larger $R_{GC}$ (e.g., Zinn 1985).  The expected change in
HB morphology due to this gradient would be for bluer HBs at larger
$R_{GC}$, the opposite of what is observed.  If age is indeed the
dominant ``second parameter'' controlling HB morphology (e.g.,
Chaboyer et al. 1996), the inner halo must be older than the outer
halo to overcome the affects of the [Fe/H] gradient.  Calibrating
their color shift with observed globular cluster ages and horizontal
branch theory, PBS estimated that the age of the halo decreases
outward by $\sim$2 Gyr over the $2 < R_{GC} < 12$ kpc range of their
observations.  They also suggested that the broader color range in the
outer halo may indicate that it formed over a more extended period
than the inner halo.
 
Suntzeff et al. (1991) interpreted similar data, concerning the [Fe/H]
distribution of halo RRL stars with $R_{GC}$, as evidence that the
inner regions of the galaxy are older than the outer regions.
 
\section{Kinematics}

\subsection{Local HB Stars}

The kinematic properties of HB stars within several kpc of the Sun
have recently been the topic of several studies involving large
numbers of stars, and are correspondingly well-determined.  The RRL
sample of Layden (1995) contains 130 stars with [Fe/H] $< -1.3$ (the
low metallicity cut isolates the halo from contaminating thick disk
stars).  Figure 2a shows the Frenk \& White (1980) analysis of this
data (i.e., based on radial velocities alone; the slope of the
least-squares fit in this plane gives the net rotation of the
population about the Galactic center, $V_{rot}$, and the dispersion
around the fit gives the line-of-sight velocity dispersion,
$\sigma_{los}$).  Layden obtained $V_{rot} = 18 \pm 13$ km~s$^{-1}$
and $\sigma_{los} = 116 \pm 7$ km~s$^{-1}$.
 
\begin{figure}
\vskip -0.4in
\plottwo{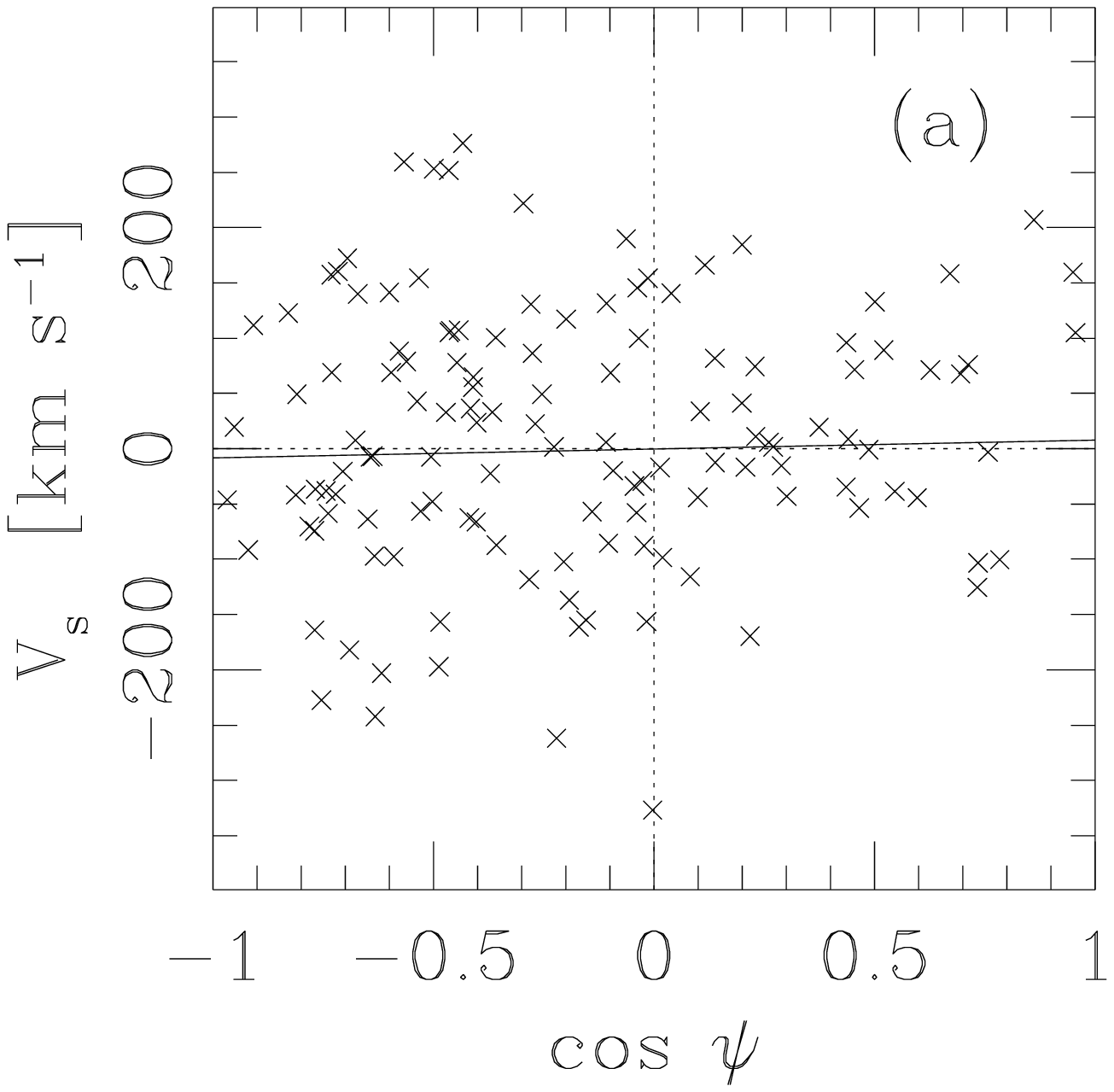}{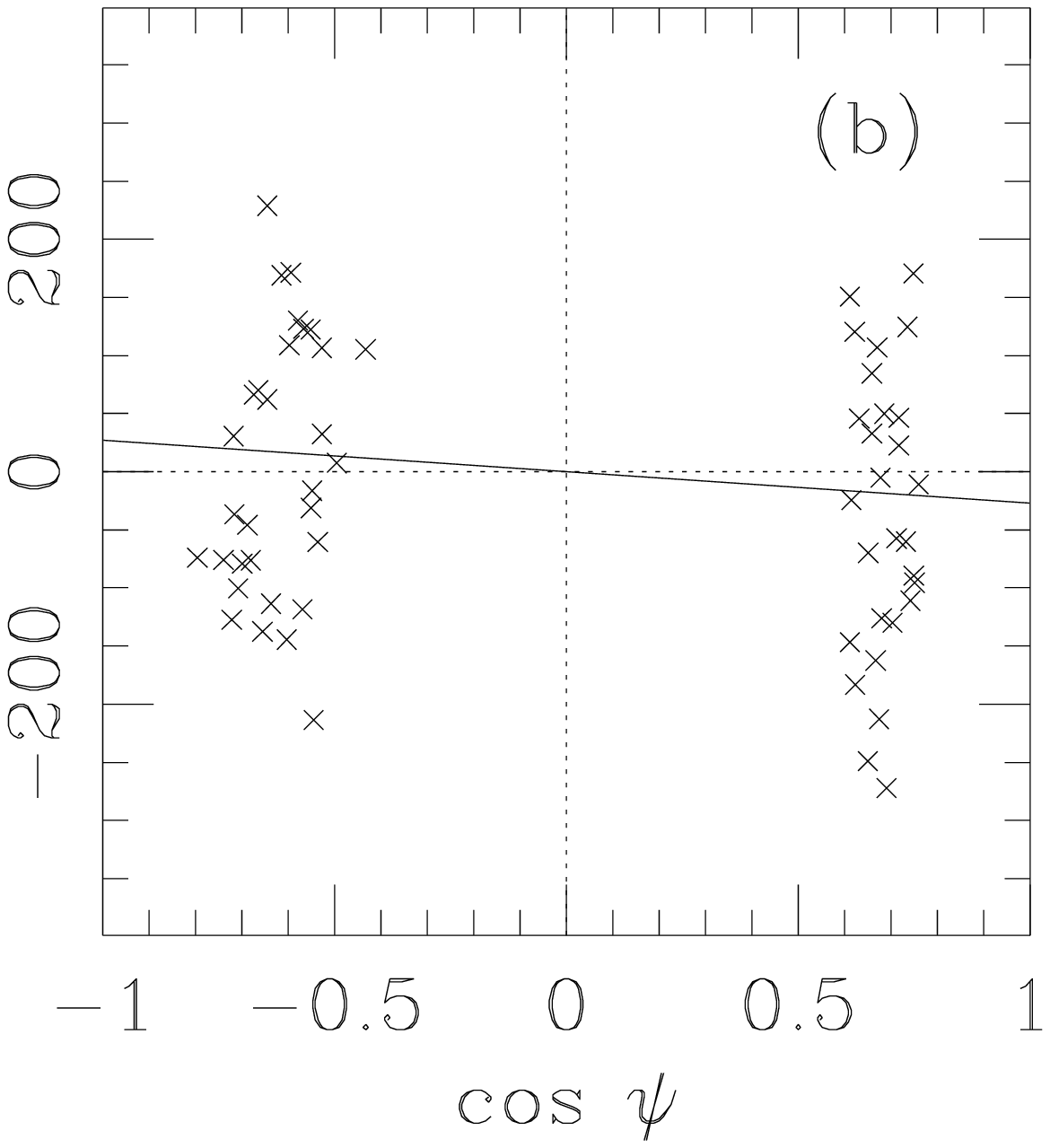}
\vskip -0.25in
\caption{Frenk \& White (1980) velocity plots for (a) the 
local RRL stars (see \S 4.1) and (b) the high halo RRL stars 
(see \S 4.2).} \label{fig-2}
\end{figure}

Meanwhile, Wilhelm et al. (1996) recently reported progress on a
large sample of BHB stars, a subset of which defines the kinematics of
the local BHB population.  Their Figure 2 shows the Frenk \& White
(1980) analysis of their data.  They obtained $V_{rot} = 40 \pm 17$
km~s$^{-1}$ and $\sigma_{los} = 103 \pm 5$ km~s$^{-1}$ for 199 BHB
stars with [Fe/H] $< -1.6$ (their data indicated an even lower
metallicity cut to isolate the halo) and $|Z| < 4$ kpc from the
Galactic plane.
 
Both samples show the slow prograde rotation and large velocity
dispersion typical of the halo, as traced by globular clusters
(Armandroff 1989), field subdwarfs (Ryan \& Norris 1991, Carney et
al. 1996), and field red giants (Morrison 1990).  Thus the local HB
star kinematics are in good agreement with other tracers of the
stellar halo.

Velocity dispersions in the three cardinal Galactic directions have
been determined for the RRL sample using proper motions and radial
velocities to get the space motions of each star (Layden 1995):
($\sigma_\rho, \sigma_\phi, \sigma_z$)= ($166\pm14, 109\pm9, 95\pm5$)
km~s$^{-1}$.  Again, these agree well with the velocity ellipsoids
of other halo tracer stars.
 
The variation of kinematics with abundance in the halo has long been
used as a diagnostic of the processes which formed the halo.  Eggen et
al.  (1962) described a formation scenario in which kinematics were
correlated with abundance, i.e., chemical enrichment took place
concurrently with the dynamical transition from spherical halo to
flattened disk.  The other extreme was promoted by Searle \& Zinn
(1978), who suggested that the halo formed via the accretion of
individual ``fragments''.  Each fragment underwent chemical evolution
in isolation from the other fragments, and later reached dynamical
equilibrium with the forming Galaxy.  Correlations between kinematics
and abundance were erased in the resulting halo.
 
The local HB stars provide a useful check on this point.  Layden
(1995) and Beers \& Sommer-Larsen (1995) have divided their samples
into small [Fe/H] bins and computed $V_{rot}$ and $\sigma_{los}$ for
each bin.
For both RRL and BHB stars, the run of
these kinematic indicators with [Fe/H] appears to be flat for
abundances below the suspected thick disk cut-off.  Thus they support
the idea that the halo stars (near the Sun) formed in proto-Galactic
fragments which were later accreted by the forming Galaxy.  The larger
samples of field subdwarfs (e.g., Ryan \& Norris 1991, Carney et
al. 1996) show this effect even more clearly.  Norris (1994) discusses
the possibility that a kinematically cooler, dissipatively formed
component might ``hide'' within the the hot velocity dispersions of
the accreted component.

\subsection{The High Halo}

The kinematic properties of stars lying at large distances from the
Galactic plane have recently provided tantalizing clues about the
structure and formation of the halo (e.g., Majewski 1992).  Samples of
HB stars in this region of the Galaxy are now becoming available for
kinematic study.
 
Wilhelm et al. (1996) reported initial results for BHB stars with
[Fe/H] $< -1.6$ and $|Z| > 4$ kpc.  Their Figure 2 shows the Frenk \&
White (1980) analysis for 90 such stars: $V_{rot} = -93 \pm 36$
km~s$^{-1}$ and $\sigma_{los} = 116 \pm 9$ km~s$^{-1}$.  Preliminary
results for RRL stars (Layden 1996) show a qualitatively similar,
though less extreme, result: $V_{rot} = -27 \pm 23$ km~s$^{-1}$ and
$\sigma_{los} = 116 \pm11$ km~s$^{-1}$ for 59 stars (see Figure 2b).
Note that though the number of stars is smaller in the Layden work,
their distribution in the (cos $\psi, V_s$) plane is better-suited to
determining $V_{rot}$.
 
In both samples, the net rotation of the halo appears to be
retrograde.  Though surprising, this result is in agreement with the
field subdwarf work of Majewski (1992) and Carney et al. (1996).
These $V_{rot}$ values assume a local circular velocity of $V_{LSR} =
220$ km~s$^{-1}$.  If $V_{LSR} = 250$ km~s$^{-1}$, the RRL rotation
becomes approximately zero, though the other surveys still indicate a
marginally retrograde high halo.  In either case, it is clear that the
rotation of the high halo is slower than that of the halo near the Sun
by 50--100 km~s$^{-1}$.  This is consistent with the concept of a
two-component halo, one of which is flattened toward the plane and
which possesses higher angular momentum.  This idea will be pursued in
the Conclusions section.
 
Another intriguing revelation concerning the high halo comes from
Kinman et al. (1996), who studied the radial velocities of RRL and BHB
stars far from the plane in the direction of the North Galactic Pole.
They found that among a group of 24 stars in one field with $4 < |Z| <
11$ kpc, the mean radial velocity was $-59 \pm 16$ km~s$^{-1}$.  In a
nearby field, the 16 stars in the same distance range had a mean
radial velocity of $-34 \pm 27$ km~s$^{-1}$.  They interpreted this as
evidence for kinematic substructure in the halo, perhaps the streaming
motions associated with a long-dissolved proto-Galactic fragment of
the type envisioned by Searle \& Zinn (1978).  Majewski et al. (1996)
found very similar vertical velocities for their NGP subdwarfs, and
noted kinematic clumping in the radial and tangential velocities as
well.  However, the velocity dispersions found by both studies are
large, a quality difficult to understand in the context of long-lived
star streams.  Perhaps instead the flow represents a transitory
``wake'' induced in the stellar halo by the tidal action of the
Magellanic Clouds or the Sagittarius dwarf (see M. Weinberg, this
volume).

\subsection{The Outer Halo}

The kinematics of stars in the outer halo ($R_{GC} > 10$ kpc) provide
further clues to the formation of the halo.  Sommer-Larsen et
al. (1997) present radial velocity data for 679 BHB stars in four
directions within the Galaxy.  They binned the stars by distance in
each field and estimated the radial velocity dispersion of the stars
in each bin out to mean distances of $d \approx 20$ kpc.
 
Since the velocity ellipsoid near the Sun is radially anisotropic
(Sec. 4.1), some variation in $\sigma_{los}$ is expected as the
contribution of the different Galactic velocity components ($\rho \phi
z$) changes with distance along each line of sight.  However, at large
$d$, all the lines of sight mainly measure the Galacto-radial ($\rho$)
velocity component, and their observed values of $\sigma_{los} =
100$--110 km~s$^{-1}$ are significantly smaller than the adopted local
value of $\sigma_\rho = 140$ km~s$^{-1}$.

To further investigate this effect, they chose a formulation for the
run of $\sigma_\rho$ with $R_{GC}$ with four adjustable parameters.
They used the Jeans equation to relate the tangential velocity
dispersion $\sigma_t(R_{GC})$ to $\sigma_\rho(R_{GC})$, thus enabling
the computation of $\sigma_{los}$ as a function of distance along any
line of sight.  Fitting this model (via the four parameters in
$\sigma_\rho(R_{GC})$) to their BHB star data (see their Figure 2),
they found that the nature of the halo velocity ellipsoid changes
dramatically outside the solar circle.  The radial velocity dispersion
drops and the tangential dispersion increases (see their Figure 1),
reaching the asymptotic (large $d$) values of 89 and 137 km~s$^{-1}$,
respectively.

They note that if the halo formed through a monolithic collapse of a
proto-Galactic cloud, one might expect the outer reaches to be
characterized by infall, i.e., radially anisotropic motions.  They
suggest that their observed tangentially anisotropic velocity
ellipsoid is more easily reconciled with fragment accretion pictures.
 
A possible complication to the arguments of Sommer-Larsen et
al. (1997) is the observed velocities of RRL stars in the inner halo
by Layden (1996).  Though preliminary, those results suggest that the
tangential velocity dispersion at $R_{GC} = 3$--6 kpc is quite large,
$\sim$150 km~s$^{-1}$, in stark contrast to the low value of $\sigma_t
= 95$ km~s$^{-1}$ predicted by the Sommer-Larsen et al. model.  It is
possible that the modeler's assumed potential field is
oversimplified, especially in the flattened inner halo.
 
\section{Conclusions}

In summary, HB star observations provide the following insights into
the structure and formation of the Galactic halo.  (1) The RRL density
distribution indicates that either the halo becomes increasingly
flattened at smaller $R_{GC}$, or that a two-component halo exists with
the flatter component increasingly dominant at smaller $R_{GC}$.  (2)
The HB morphology of the halo becomes redder, and perhaps broader,
with increasing $R_{GC}$; if interpreted as an age effect, the outer
halo is several Gyr younger, in the mean, than the inner halo, and may
have formed over a longer period of time.  (3) Evidence exists that at
least a part of the halo at $R_{GC} \geq R_\odot$ formed via accretion
of proto-Galactic fragments: (a) lack of a kinematics--metallicity
correlation in local HB star samples, (b) slower/retrograde
rotation of the high halo, (c) possible presence of star streams
in the high halo, and (d) transition from radially to tangentially
anisotropic velocity dispersions in the outer halo.  (4) The RRL
density distribution and the difference between the rotation rates of
the local and high halo samples both support the notion of a
two-component halo.

These HB star observations fit into the larger picture of Galactic
halo formation outlined by Zinn (1993, 1996) and expanded upon by
Norris (1994) and Carney et al. (1996).  Specifically, two distinct
halo components exist, which were created through different
mechanisms.  One component, variously known as the ``old'',
``flattened'', or ``proto-disk'' component, has a significant prograde
rotation, a flattened spatial distribution, and a relatively blue HB
morphology.  It may possess a metallicity gradient, and appears to
dominate at $R_{GC} \leq R_\odot$ and near the Plane.  Its properties
suggest formation through a fairly ordered process of collapse and
spin-up, not unlike that described by Eggen et al. (1962).  The other
component, referred to as ``younger'', ``spherical'', or ``accreted'',
has a retrograde net rotation, a spherical spatial distribution, and
shows no evidence of a metallicity gradient.  Organized streaming
motions may be found among the stars in this component, relics of the
small galaxies that were accreted during its formation.  Cluster ages
and HB morphologies suggest that it formed over a period of many Gyr,
and is on average several Gyr younger than the other halo component
(e.g., Chaboyer et al. 1996).  These properties are all signposts of
the fragment accretion picture of Searle \& Zinn (1978).  These two
halo components have very similar metallicity distributions, and 
are difficult to separate in local stellar samples.  Thus, the current
halo formation picture may be regarded, at least in a first
approximation, as a fusion of the classical ideas of Eggen et
al. (1962) and Searle \& Zinn (1978).

\acknowledgments 

Support for this work was provided by NASA through grant number
HF-01082.01-96A from the Space Telescope Science Institute, which is
operated by the Association of Universities for Research in Astronomy,
Inc., under NASA contract NAS5-26555.



\begin{references}

Armandroff, T.E. 1989, \aj, 97, 375

Beers, T.C. \& Sommer-Larsen, J. 1995, \apjs, 96, 175

Carney, B.W., Laird, J.B., Latham, D.W., \& Aguilar, L.A. 1996, \aj,
112, 668

Chaboyer, B., Demarque, P., \& Sarajedini, A. 1996, \apj, 459, 558

Eggen, O.J., Lynden-Bell, D., \& Sandage, A.R. 1962, \apj, 136, 748

Frenk, C.S. \& White, S.D.M 1980, \mnras, 193, 295

Hawkins, M.R.S. 1984, \mnras, 206, 433

Kinman, T.D. 1965, \apjs, 11, 199

Kinman, T.D., Suntzeff, N.B., \& Kraft, R.P. 1994, \aj, 108, 1722

Kinman, T.D., et al. 1996, \aj, 111, 1164

Layden, A.C. 1995, \aj, 110, 2288

Layden, A.C. 1996, ASP Conf. Ser. 92, 141

Lee, Y.-W., Demarque, P., \& Zinn, R. 1994, \apj, 423, 248

Majewski, S.R. 1992, \apjs, 78, 87

Majewski, S.R., Munn, J.A., \& Hawley, S.L. 1996, \apj, 459, L73

Minniti, D. 1996, \apj, 459, 175

Morrison, H.L., Flynn, C.M., \& Freeman, K.C. 1990, \aj, 100, 1191

Preston, G.W., Shectman, S.A., \& Beers, T.C. 1991, \apj, 375, 121
(PSB)

Ryan, S.G. \& Norris, J.E. 1991, \aj, 101, 1835

Searle, L. \& Zinn, R. 1978, \apj, 225, 357


Sommer-Larsen, J., et al. 1997, \apj, 481, 775

Suntzeff, N.B., Kinman, T.D., \& Kraft, R.P. 1991, \apj, 367, 528

Wetterer, C.J. \& McGraw, J.T. 1996, \aj, 112, 1046


Wilhelm, R., et al. 1996, ASP Conf. Ser. 92, 171


Zinn, R. 1985, \apj, 293, 424



Zinn, R.J. 1996, ASP Conf. Ser. 92, 211

\end{references}
\end{document}